\documentclass[10pt, a4paper, twocolumn, copyright, goog]{google}

\usepackage{listings,xcolor}
\usepackage{color,amsmath,enumitem}
\usepackage{language_formatting}
\usepackage{graphicx}
\usepackage[authoryear, sort&compress, round]{natbib}
\usepackage{xurl}

\title{Toward provably private analytics and insights into GenAI use}

\correspondingauthor{dramage@google.com}

\author[*, 1]{Albert Cheu}
\author[*, 1]{Artem Lagzdin}
\author[1]{Brett McLarnon}
\author[1]{Daniel Ramage}
\author[*, 1]{Katharine Daly}
\author[1]{Marco Gruteser}
\author[1]{Peter Kairouz}
\author[1]{Rakshita Tandon}
\author[1]{Stanislav Chiknavaryan}
\author[*, 1]{Timon Van Overveldt}
\author[1]{Zoe Gong}

\affil[*]{Primary authors of this paper}
\affil[1]{\thepa{}{}}

\begin{abstract}
  Large-scale systems that compute analytics over a fleet of devices must achieve high privacy and security standards while also meeting data quality, usability, and resource efficiency expectations. We present a next-generation federated analytics system that uses Trusted Execution Environments (TEEs) based on technologies like AMD SEV-SNP \citep{amdsevsnp} and Intel TDX \citep{intelTDX} to provide verifiable privacy guarantees for all server-side processing. In our system, devices encrypt and upload data, tagging it with a limited set of allowable server-side processing steps. An open source, TEE-hosted key management service guarantees that the data is accessible only to those steps, which are themselves protected by TEE confidentiality and integrity assurance guarantees. The system is designed for flexible workloads, including processing unstructured data with LLMs (for structured summarization) before aggregation into differentially private insights (with automatic parameter tuning). The transparency properties of our system allow any external party to verify that all raw and derived data is processed in TEEs, protecting it from inspection by the system operator, and that differential privacy is applied to all released results. This system has been successfully deployed in production, providing helpful insights into real-world GenAI experiences.
\end{abstract}

\begin{document}

\maketitle

\section{Introduction}

Generative AI enables personalized experiences and powers the creation of unstructured data including summaries, transcriptions, and more. Understanding how users engage with powerful GenAI tools can provide insights into real-world AI usage \citep{clio2024, chatgpt2025} and help GenAI developers make their tools even more useful by understanding common patterns and failure modes. However, the process of deriving meaningful insights from user data presents serious privacy challenges, especially when applied to sensitive or on-device data. This paper describes the challenge of offering ``provably private'' analytics and insights on unstructured data and describes a system that sets a new bar in terms of its formal privacy guarantees. Extending earlier work in federated analytics \citep{google_federated_analytics_2020, eichner2024confidential}, our system leverages the power of large language models (LLMs), differentially private (DP) aggregation, and trusted execution environments (TEEs) to generate dynamic, provably private insights into how people use LLMs and GenAI tools.

To build a provably private system for analytics and insights, we must first define the set of parties involved and their goals, the set of privacy threats to be mitigated, and what constitutes a standard of ``proof.'' In our setting, a service provider deploys AI tools to a user population. These tools may operate on unstructured inputs and outputs, like speech, text, images, or video. The goal of a privacy-preserving analytics system is to enable the service provider to gain aggregate insights into the tool's usage while (a) observing no unaggregated data or intermediates and (b) observing only aggregated output that has been generated with an appropriate formal anonymization guarantee such as differential privacy (DP) \citep{DMNS06}. We assume that the service provider, if they are the operator of the analytics system, may attempt to observe or even interfere with processing in order to learn facts about an individual. Our provably private system for analytics and insights offers external verifiers the ability to tell if properties (a) or (b) are violated\footnote{While our system prevents direct access to unaggregated data or intermediates, it does not yet protect against possible data leakage through side-channels.}, even without the service provider's cooperation in the verification process.

Our system achieves this by using Trusted Execution Environments (TEEs) backed by technologies like AMD SEV-SNP \citep{amdsevsnp} and Intel TDX \citep{intelTDX}.
\footnote{Our system assumes that the TEE hardware manufacturer and analytics service provider are independent and non-colluding.}
TEEs provide confidentiality through memory encryption and integrity of executed code, and they allow that code to be remotely attested. Our system leverages these capabilities to enable devices to independently upload data with the guarantee that subsequent privacy-preserving server-side processing logic will run prior to any derived (aggregate) data becoming inspectable to the service provider. Service providers who use our system configure both an on-device query that is deployed to devices as well as a server-side workflow that supports LLM inference and aggregation with parameters that are automatically tuned to achieve high-utility central DP guarantees. A DP algorithm mathematically bounds the ability to re-identify or infer the data belonging to a given DP unit, which in our current system is a single device upload, from the output.

The next section of this paper discusses related work on large-scale analytics systems and other systems involving TEEs. Following that, we describe the components of our system and detail how we achieve verifiable privacy guarantees given our trust model. After establishing this baseline description of our system, we go into more detail about certain advanced features, including LLM inference and DP autotuning. Lastly we discuss future improvements we plan to make to our system, including supporting accelerators, flexible DP units, and user-level $k$-anonymity.

\section{Related Work}

Large-scale systems that compute analytics over a fleet of devices have been used to enable improvements in on-device language models used by the Android keyboard \citep{sun2024private}, identify iconic scenes in Apple Photos \citep{apple2023}, and provide insights into aggregate health experiences via Google Health Studies \citep{ghs2020}. Given the sensitivity of on-device data, the privacy expectations of on-device data owners, and the risk exposure for companies with access to on-device data, privacy and security guarantees are an important part of these analytics systems. Different systems have addressed privacy and security goals in a variety of ways.

Systems that use local DP \citep{apple2017} apply noise to client uploads such that the server only sees data that is already private, but the utility of the final results suffers in comparison to central DP methods where noise is applied on the server post-aggregation. Multi-party computation (MPC) systems utilize secret sharing amongst clients to ensure that the service provider is only able to interpret the data after aggregating uploads from many clients \citep{bonawitz2017practical}, or they split devices’ data among multiple servers assumed to be operated by non-colluding entities to ensure no single server is able to view one device’s raw data \citep{corrigan2017prio}. In addition to requiring advanced coordination, MPC systems are susceptible to Sybil attacks, prompting some MPC systems to also incorporate distributed DP schemes \citep{kairouz2021distributed, talwar2024samplable} where a small amount of noise is added to client uploads such that aggregating these noised client uploads produces a final noised result mirroring a central DP guarantee.

Trusted execution environments (TEEs) have also been used in recent systems requiring externally verifiable guarantees for server-side processing of sensitive data due to their ability to provide data confidentiality and workload integrity. With modern models becoming too large to run on-device, TEEs have been deployed server-side to run LLM inference over sensitive data \citep{apple2024, meta2025}, with the privacy narratives of these deployments differing on the level of external inspectability and operator trust required. Papaya supports large-scale analytics \citep{srinivas2025papaya} and learning \citep{huba2022papaya} via separate stacks that both use TEEs server-side, and this paper describes improvements to an earlier iteration of our multi-purpose TEE-based solution \citep{eichner2024confidential}. There are subtle differences between our system and Papaya in how we achieve large-scale private analytics processing via TEEs, including which types of server-side workloads are supported, how server-side work is scheduled and executed, and how verifiable privacy is provided.

\section{System Description}
\label{system_description}
\newcommand{\eps}{\varepsilon}

\begin{figure*}[h!]
    \centering
    \includegraphics[width=0.9\textwidth]{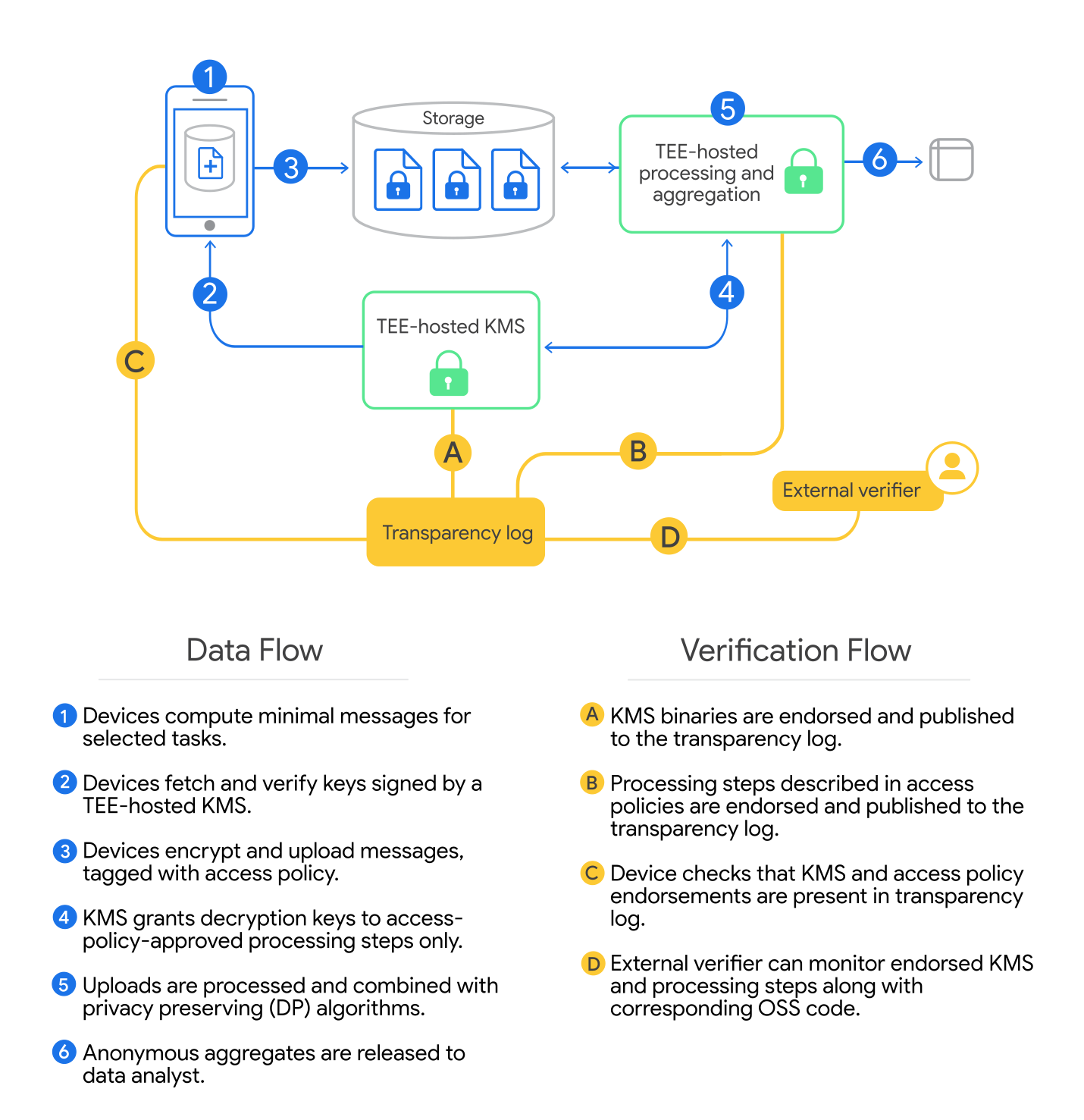}
    \caption{The data and verification flow in our TEE-based private analytics system.}
    \label{fig:cfa}
\end{figure*}

Our data processing system, shown in Figure \ref{fig:cfa}, consists of three main components:
\begin{itemize}
\item A \emph{key management service (KMS)}, which runs in a TEE and which manages keys for encrypting private data and processing that data using publicly verifiable pipelines.
\item A \emph{client library} for verifying KMS-managed public encryption keys before upload, running on end-user devices.
\item \emph{Data processing pipelines}, in which untrusted orchestration logic routes encrypted data amongst a collection of data processing binaries running in TEEs, with the KMS mediating all access to decryption and encryption keys, in order to produce a final, verifiably private result.
\end{itemize}

These components are central to achieving the following system properties:
\begin{itemize}
\item \textbf{Verifiability of data processing.} Devices must be able to know prior to uploading data which server-side processing binaries will have access to uploaded data and under what constraints (e.g. differential privacy parameters) they will be able to operate over the uploaded data. Our client library tags uploaded data with \emph{data access policies} to capture this information.
\item \textbf{Confidentiality.} All server components that access unencrypted client data must run in a TEE and be remotely attestable. The KMS only provides decryption keys to binaries that match the access policy attached to uploaded data.
\item \textbf{Asynchrony.} The client device to processing pipeline communication must be secure, compatible with standard upload mechanisms (HTTP POST to temporary storage), and asynchronous. The KMS acts as a trusted mediator between the client devices and the server-side processing binaries, allowing the client upload stage to be decoupled from the server-side processing stage.
\item \textbf{Best-effort TTL enforcement.} The uploaded data must be subject to a time-to-live (TTL), although enforcement of this TTL is not verifiable.
\item \textbf{Scalability.} The system should scale to billions of records uploaded per day and have high availability.
\end{itemize}

In the following subsections we describe the KMS and access policies in more detail, discuss how the privacy properties of our system can be verified by external parties, and introduce a canonical server-side workload supported by our system.

\subsection{Key Management Service}
Conceptually, the Key Management Service (KMS) is a fault-tolerant service which combines a key store for holding keys needed to decrypt uploaded data with a small stateful database for tracking which inputs have already been processed. Combining the two services ensures that the usage database cannot be cleared or disrupted without also affecting the corresponding decryption keys.

The KMS enables secure asynchronous communication between data producers and consumers using pre-existing channels, such as uploads with HTTP POST, transmission with Google Cloud Pub/Sub, or processing with Apache Beam. Client devices and non-terminal processing steps are data producers; processing steps are data consumers.

\subsubsection{Design Overview}
The KMS is implemented as a TEE-based key management service that issues hybrid public key encryption (HPKE) keys for protecting both uploaded data and short-lived intermediate files produced by processing pipelines. These pipelines are also TEE-based, and the KMS will only share decryption keys with the pipeline worker nodes running the specific software versions authorized by an access policy selected at the time the data was uploaded. Client devices are therefore able to know at upload time precisely in which ways their data is permitted to be used.
Client devices require that both the access policy and KMS software have been recorded in a public transparency log, preventing split-view attacks where different clients are served different access policies or KMS implementations.

The KMS also stores rollback-protected state for each pipeline, allowing the pipelines to track privacy budgets using a format determined by that pipeline. These budgets often have very concise representations that do not scale with the number of uploads (e.g. ``all data before 2025-01-01 has no remaining budget''); in cases where the budget-tracking state is large, pipelines can instead store a cryptographic checksum. State updates use optimistic concurrency: pipelines are allowed to run and produce encrypted outputs without updating the state, but the pipelines can make a state mutation a prerequisite for releasing an output outside of a TEE. Note that the KMS guarantees integrity but not secrecy for the pipeline state.

Encryption keys and pipeline state are stored in the same replicated key-value database that is kept in-memory in TEEs and replicated using the RAFT consensus protocol \citep{raftusenix}. Any attempt to clear stored pipeline state by restarting the RAFT cluster would also destroy the decryption keys for uploaded data and pipeline intermediates.

\subsubsection{Access Policies}
An access policy describes which data processing pipelines can access uploaded data, the specific binaries making up those pipelines, and any constraints associated with the processing logic performed by the binaries (e.g. differential privacy parameters).

An access policy is composed of three different layers:
\begin{itemize}
   \item The top-level policy specifies the different data processing pipelines that are authorized to access the data. Specifically, it contains one or more logical pipeline policies. 
   \item A logical pipeline policy describes one or more authorized variants (i.e. versions) of a pipeline. Supporting multiple versions enables access policy updates to be decoupled from rolling out new processing pipeline versions. While the logical pipeline policy can contain multiple variants, usually only one of those variants will actually process the uploaded data.
   \item A pipeline variant policy describes the specific sequence of TEE-hosted binaries that are authorized to process the uploaded data, as well as any privacy-relevant configuration for those binaries (e.g. differential privacy parameters). They encode a directed bipartite graph, one set of nodes representing data in the system (e.g. initial uploads, pipeline intermediates) and the other nodes representing data processing steps that consume inputs and produce outputs.
\end{itemize}

The use of access policies allows a client to delegate remote attestation responsibilities to the KMS: clients only attest a single server (the KMS), not a whole chain of them. This (a) reduces the number of attestations sent to the client, saving bandwidth, and (b) allows the processing jobs to be started after the client uploads data, simplifying server management.

\subsubsection{TTLs via Crypto-Erasure}

Upload encryption keys generated by the KMS have expiration times, after which the KMS erases them from its memory, preventing further use. Encryption keys for pipeline intermediates similarly have expiration times, and when a pipeline requests decryption keys for its inputs from the KMS, it will not receive any keys that expire before the encryption keys for its outputs. This ensures that a well-behaved pipeline doesn't accidentally extend the lifespan of input data beyond the intended TTL.
However, getting access to a trusted time source is difficult — especially at high QPS. Instead, the KMS maintains a replicated, monotonically increasing record of the observed system time; this time is used to (a) set issued at and expiration times on new keys and (b) trigger the erasure of expired old keys.

If the (untrusted) system time is far in the past, the keys generated by the KMS will already be expired and clients will not upload data.
If the (untrusted) system time is far in the future, keys generated will not be valid yet and clients will not upload data. Moreover, the KMS will expire any data that was previously uploaded with a more accurate timestamp.
The system operator is therefore motivated to provide an accurate clock.

\subsection{External Verifiability}

Clients require that the public encryption key they encrypt their data with was produced by a KMS whose software, running in a TEE, is published to a public transparency log.
They do this by verifying a transparency log inclusion proof of the KMS's software and TEE attestation evidence, and by verifying that the encryption key is rooted, via a signature chain, in that inclusion proof.
Clients also require that data access policies are published to a transparency log, by similarly validating an inclusion proof for the digest of the access policy.
Both the KMS software and data processing software are reproducibly buildable from open-source code.
This ensures that both the KMS software and data access policies are auditable: anyone can monitor the transparency log entries and trace the KMS software binary digests and access policy digests to the corresponding KMS and data processing source code.

Server-side pipelines consist of untrusted orchestration logic that passes encrypted data amongst TEEs running data processing binaries until a persistent state update request is finally sent to the KMS in exchange for a decryption key allowing access to a final result. Because access policies specify not just the data processing binaries but also their configuration, the process by which client data is transformed by the processing TEEs into a final result also becomes verifiable, even in the event of a malicious orchestrator. While the KMS only provides decryption keys to data processing binaries specified in the access policy, it does not preclude attacks that inject Sybil inputs or leverage side-channel leakage. Side-channels include lengths of encrypted data, timing of TEE-hosted applications, and memory access patterns by those applications. As discussed in our prior work \citep{eichner2024confidential}, verifiable DP addresses Sybils but mitigating side-channels is an ongoing research and engineering effort.

\subsection{Differentially Private Aggregation}
The canonical use case for our system thus far in Google production use has centered around running server-side SQL group-by aggregations over uploaded client data with verifiable DP guarantees. We support both a closed-set DP variant (where the domain of possible groups is specified alongside the group-by query) and an open-set DP variant (where the domain is not known in advance). In addition to configuring an upload task and a server-side aggregation SQL query, users of our system have the option to specify additional server-side per-DP-unit work that runs prior to the aggregation SQL query. The server-side per-DP-unit work could be a SQL query that will run on each DP unit independently, or as will be discussed further in Section \ref{inference}, it could involve running LLM inference over each DP unit independently. Currently we treat each upload as a separate DP unit, but in Section \ref{flexible_dp_units}, we will describe future plans to increase the flexibility of our system by allowing uploads to consist of multiple DP units and/or partial DP units.

In order to provide a verifiable DP guarantee for this canonical use case, we must prove (a) any server-side per-DP-unit work is applied at precisely the DP unit scope, (b) a ($\varepsilon$, $\delta$)-DP aggregation algorithm is executed that contribution bounds each DP unit, aggregates, then noises/thresholds the results, and (c) the privacy budget in the pipeline state tracks the appropriate information. The access policy we use for these workloads ensures all of this by specifying the required data processing binary, the required runtime configuration for the data processing binary (this includes the $\varepsilon$ and $\delta$ values as well as an identifier for the DP aggregation algorithm that must run), and a limit on how many times a given client upload may be used to produce an aggregated result. Notably, in order to provide a verifiable DP guarantee we do not actually need to prove the exact server-side per-DP-unit pre-aggregation work that is applied nor the exact DP unit contribution bounds that are used, which is why these values are not included in the access policy. See Appendix \ref{app:access_policy} for an example of this access policy. The pipeline maintains KMS-protected state to ensure that client uploads are included in no more than the configured maximum number of aggregate results.

\section{Structured Summarization}
\label{inference}
Understanding how users engage with the diverse features of modern GenAI tools yields valuable insights and identifies failure modes. Developers can then use this data to progressively enhance the user experience \citep{clio2024, chatgpt2025}. However, analyzing GenAI usage requires processing sensitive user data, particularly with on-device model deployments. This data is often unstructured, making traditional SQL-based analytics methods insufficient for processing it.

Consider classifying text transcripts stored on a device. The traditional method of sending them to a server for processing creates a significant privacy risk by exposing the raw data. An alternative is processing data directly on-device, but the quality of this approach is capped by the limitations of smaller on-device models, and this approach can also place a heavy computational load on the device, degrading the user experience.

To address these concerns, we deploy LLM-based \textit{structured summarization} inside a TEE. This method allows for the intelligent processing of sensitive user data while guaranteeing the raw data is never exposed. The entire process is confined to the TEE, ensuring that only aggregated and anonymized results are available for analysis and preserving the privacy of the original data.

We implement this system by integrating \mbox{gemma.cpp}, a lightweight C++ inference engine for the Gemma foundation models \citep{gemmacpp}, into TEE-hosted binaries \citep{cfc}.  To enforce strict data isolation, each inference run processes only a single upload's data, and the engine's internal state is completely re-initialized afterward to prevent cross-upload data contamination. This inference step is performed before the server-side per-DP-unit SQL query and DP aggregation. Even if a prompt attempts to single out a single device upload, the DP guarantee ensures that the information from this upload will not be revealed.

\begin{figure*}[h]
    \centering
    \includegraphics[width=0.9\textwidth]{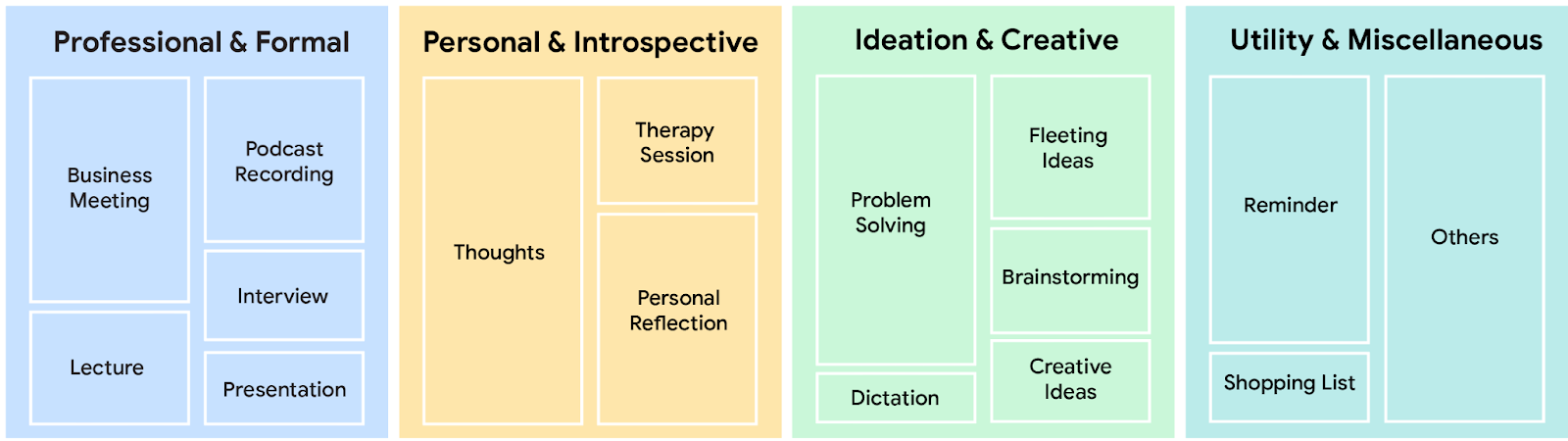}
    \caption{An example differentially private distribution of transcripts in the Recorder application across various topics, as categorized by the Gemma model. Inner rectangle size is proportional to relative topic frequency.}
    \label{fig:recorder}
\end{figure*}

The system is compatible with any open-source Gemma model (\cite{kagglegemma}). The model weights and the associated prompt are provisioned at runtime through an API exposed by the TEE-hosted binary. We have successfully used Gemma3 4B models in our production deployments using AMD Secure Encrypted Virtualization-Secure Nested Paging (SEV-SNP) CPUs \cite{amdsevsnp}. Figure \ref{fig:recorder} shows an example DP distribution of transcript topics within the Google Recorder application \cite{recorder}, as categorized by the Gemma model.

\section{Contribution Bounds and Autotuning}

\paragraph{Initial queries.} In the initial version of our system \cite{eichner2024confidential}, we processed server-side workloads with DP aggregation logic taking the following syntax:

\begin{lstlisting}[language=googlesql]
SELECT
    WITH DIFFERENTIAL_PRIVACY OPTIONS(epsilon=[number], delta=[number], max_groups_contributed=[number])
    key_1,
    key_2,
    ... ,
    SUM(value_1) @{L_inf=[number], L_1=[number], L_2=[number]} AS total_value_1,
    SUM(value_2) @{L_inf=[number], L_1=[number], L_2=[number]} AS total_value_2,
    ...
FROM ClientQueryResults
GROUP BY key_1, key_2, ...
\end{lstlisting}

The parameters \texttt{max\_groups\_contributed, L\_inf, L\_1}, and \texttt{L\_2} constitute \emph{contribution bounds}, which control how much any single DP unit can influence the computation. To illustrate each parameter's meaning, consider a query that adds up dollars spent across many locations. Specifically, there is one grouping column encoding locations and one numerical value being summed that tracks expenditures.
\begin{itemize}
    \item \texttt{max\_groups\_contributed}: limits the number of locations receiving money from one DP unit.
    \item \texttt{L\_inf}: limits how much one DP unit spends at any location.
    \item \texttt{L\_1}: limits the total dollars spent by one DP unit, across all locations.
    \item \texttt{L\_2}: limits the square-root of the sum of squared dollars spent, across all locations.
\end{itemize}

Differentially private algorithms each require\footnote{If the domain of the grouping keys is unknown, \texttt{max\_groups\_contributed \& L\_inf} are required to privately identify the subset of the domain present in the data. If the domain is known, either \texttt{L\_1}, \texttt{L\_2} or the pair \texttt{max\_groups\_contributed \& L\_inf} are required for the Laplace or Gaussian mechanisms.} some combination of these bounds: if we did not perform contribution bounding, there would be no finite scale of variance that would mask any DP unit's contribution. Conversely, we can prove DP guarantees for an algorithm that enforces an upper limit on contributions and scales noise linearly with that bound.

\paragraph{Challenges.} A data analyst hoping to use our system may not know these contribution bounds; the goal of analytics is, after all, to learn about an unknown dataset. In an ideal interface, the privacy parameters $\varepsilon,\delta$ would be the only necessary parameters; if the analyst knew tight contribution bounds, they could provide them as optional inputs.

Although arbitrary contribution bounds could be entered with no effect on differential privacy, doing so would almost certainly harm accuracy: overly small values will dampen the signal, leading to overly biased aggregates, while overly large values will inflate the variance beyond what is necessary.

Another approach is to apply a quantile-finding algorithm \citep{durfee2023unbounded}. For example, \texttt{max\_groups\_contributed} can be set to, say, the 83rd percentile of the number of groups that a DP unit contributes to. Previous iterations of federated analytics placed the responsibility of computing such queries on the data analyst. This a point of friction for on-boarding so repeating that design pattern risks dissuading potential users from engaging with our system (and possibly DP in general). Additionally, the sensitive data is walled off from the analyst by design, so they would have to run the quantile query on a suitable proxy dataset; acquiring public data (or generating synthetic data) that aligns with the sensitive data has its own host of challenges.

\paragraph{Our Solution.} In the latest iteration of our system, we automatically tune the parameters in question by running the quantile queries on a fraction of the data available to us. At a high level, we do the following:
\begin{enumerate}
    \item Using the SQL query, identify which contribution bounds are required but absent; these will be tuned. 
    \item Based on privacy budget $\varepsilon$ and the number of parameters to tune, randomly divide the $n$ inputs into two partitions $S,\overline{S}$; we elaborate on criteria for $S$ later.
    \item For each parameter to tune, apply a differentially private quantile algorithm on $S$ to obtain an estimate of the $k=83$rd percentile.\footnote{$k=83$ was found to result in a local minima of sample complexity when there is 1 parameter to tune; optimization for larger number of parameters is an interesting line of future work.} The $\varepsilon$ budget is split across the parameters. Though it may be tempting to use a non-DP algorithm, doing so opens the door to Sybil attacks; see Appendix \ref{app:autotuning}.
    \item Apply an $(\varepsilon,\delta)$-differentially private histogram algorithm on $\overline{S}$ using the tuned parameters from the previous step. Both the histogram and the tuned values are revealed to the user.
\end{enumerate}

$S$ is created by generating a Bernoulli sample for each upload. The Bernoulli parameter is big enough such that the outputs of the auto-tuning algorithm are 90\% likely to be representative of the whole population. For example, when tuning \texttt{max\_groups\_contributed \& L\_inf}, there is a 90\% chance of getting (a) a number between the $(k-10)$th and $(k+10)$th percentiles of the number of groups that a DP unit contributes to \& (b) a number between the $(k-10)$th and $(k+10)$th percentiles of the maximum value of a contribution. We defer mathematical formalism and analysis to Appendix \ref{app:autotuning}

The autotuning step constitutes a data processing pipeline that requires authorization from the KMS. As a result, the access policy needs to permit a bundle of DP quantile queries in addition to the DP group-by aggregation. Even when autotuning occurs prior to aggregation, the privacy parameters of the server-side workflow remain the $(\varepsilon, \delta)$ specified in the SQL aggregation query by virtue of partitioning the dataset: no DP unit is ingested by both autotuning and aggregation.

\section{Future Work}
\label{future}
\subsection{Accelerators and GenAI Applications}
Structured summarization is a foundational step toward our goal of delivering provably private insights for Generative AI features. To scale structured summarization and support LLMs exceeding 20B parameters, we are integrating our system with Intel Trust Domain Extensions (TDX) (\cite{intelTDX}) and NVIDIA H100 GPUs \citep{nvidiah100}. Leveraging the GPU confidential computing capabilities will allow us to perform the LLM inference step directly on the GPU, ensuring sensitive user data never leaves the trusted environment and greatly improving the performance and scalability properties of our privacy-preserving analytics system. This advancement will unlock new capabilities, including differentially private clustering \citep{liu2025uraniadifferentiallyprivateinsights} and synthetic data generation \citep{augenstein2020generativemodelseffectiveml, hu2025actgarldifferentiallyprivateconditional}, and more.

\subsection{Flexible DP units}
\label{flexible_dp_units}
As mentioned in Section \ref{system_description}, our system aims to decouple the collection of client data from the processing of this data server-side. Our current DP data processing pipelines treat each upload as a separate DP unit, meaning that if a customer desires a per-user / per-week DP unit, they must pay careful attention to the schedule at which devices upload data. We aspire to decouple client data collection from the concept of DP units, such that devices can upload data on an arbitrary schedule without impacting the server-side result. In some cases customers may require a TTL for on-device data that is shorter than the desired DP time unit (e.g. a 2-day on-device TTL vs a 7-day DP time unit), necessitating a different approach than simply treating each upload as a distinct DP unit.

To address these limitations, we are adding the ability to reconstruct DP units from client uploads during the server-side processing step. Customers will define the desired DP unit in the access policy by specifying a DP time unit schedule and any relevant partitions (e.g. if a customer desired per-user / per-week / per-location DP, they would specify a weekly DP time unit schedule and “location” as the relevant partition in the access policy). In this approach, uploads will not be expected to correspond to DP units in any way; uploads that contain data from multiple DP units or represent incomplete DP units are all acceptable. We will require, however, that uploads contain a privacy ID (an ID that is expected to remain consistent for at least an entire DP unit) as well as event time information to allow us to reconstruct DP units server-side. The privacy ID will be stored within the encrypted part of the upload and will thus not be visible to untrusted server code. Our client library will also support periodically rotating this ID.

Before DP units are aggregated together using a DP algorithm, any server-side mapping transformations over client data must not (a) allow the mapping output for one DP unit's data to be influenced by data belonging to a separate DP unit, or (b) re-map data to a different DP unit. Transformations satisfying these requirements may occur either before or after DP unit reconstruction, which is the process of regrouping rows of data into DP units by inspecting the privacy id, event time, and partition information for each row of data. In order to properly parallelize DP unit reconstruction across many machines, we will ensure that all data belonging to a single DP unit gets sent to the same machine. The DP aggregation that follows will apply contribution bounding at the scope of these reconstructed DP units.

\subsection{$k$-anonymity}
We are also planning to add $k$-anonymity support to our system. If a $k$ value is configured via a SQL hint in the server-side workload, data for a given partition will only be released if at least $k$ devices contributed to that partition. Unlike DP, server-side $k$-anonymity is susceptible to Sybil attacks because fabricated inputs can boost the apparent size of a group. Thus $k$-anonymity is only appropriate in cases where customers are willing to assume a trusted orchestrator (note that our default threat model assumes an untrusted orchestrator). In order to properly identify the number of unique devices contributing to a given partition during an aggregation window, customers must ensure that the privacy ID in a device’s uploads remains consistent for data belonging to that aggregation window.

\section{Conclusion}
We have described our TEE-based private analytics system, which achieves verifiable privacy guarantees under a threat model involving an untrusted server operator while also prioritizing usability and scalability. We support multiple forms of server-side workloads, including ones with a structured summarization step and/or SQL query, and we also support automatic tuning of DP parameters to enable easy onboarding for a wide range of use cases. The KMS and access policy design allows for decoupling of the client upload step and server-side pipeline step while still supporting highly scalable server-side processing. Our future plans include scaling the structured summarization capabilities by running on GPUs, improving operability by decoupling client uploads from DP units, and supporting $k$-anonymity.

\section{Acknowledgements}
The authors thank Brendan McMahan for thorough feedback on this paper.

\bibliography{main}

\begin{thebibliography}{29}
\providecommand{\natexlab}[1]{#1}
\providecommand{\url}[1]{\texttt{#1}}
\expandafter\ifx\csname urlstyle\endcsname\relax
  \providecommand{\doi}[1]{doi: #1}\else
  \providecommand{\doi}{doi: \begingroup \urlstyle{rm}\Url}\fi

\bibitem[{Advanced Micro Devices (AMD)}(2020)]{amdsevsnp}
{Advanced Micro Devices (AMD)}.
\newblock {AMD SEV-SNP: Strengthening VM Isolation with Integrity Protection and More}, January 2020.
\newblock URL \url{https://docs.amd.com/v/u/en-US/SEV-SNP-strengthening-vm-isolation-with-integrity-protection-and-more}.
\newblock Technical Report.

\bibitem[Apple(2023)]{apple2023}
Apple.
\newblock Learning iconic scenes with differential privacy.
\newblock \url{https://machinelearning.apple.com/research/scenes-differential-privacy}, 2023.

\bibitem[Augenstein et~al.(2020)Augenstein, McMahan, Ramage, Ramaswamy, Kairouz, Chen, Mathews, and y~Arcas]{augenstein2020generativemodelseffectiveml}
S.~Augenstein, H.~B. McMahan, D.~Ramage, S.~Ramaswamy, P.~Kairouz, M.~Chen, R.~Mathews, and B.~A. y~Arcas.
\newblock {Generative Models for Effective ML on Private, Decentralized Datasets}, 2020.
\newblock URL \url{https://arxiv.org/abs/1911.06679}.

\bibitem[Bonawitz et~al.(2017)Bonawitz, Ivanov, Kreuter, Marcedone, McMahan, Patel, Ramage, Segal, and Seth]{bonawitz2017practical}
K.~Bonawitz, V.~Ivanov, B.~Kreuter, A.~Marcedone, H.~B. McMahan, S.~Patel, D.~Ramage, A.~Segal, and K.~Seth.
\newblock Practical secure aggregation for privacy-preserving machine learning.
\newblock In \emph{ACM SIGSAC Conf. on Comp. and Comm. Security}, pages 1175--1191. ACM, 2017.

\bibitem[Chatterji et~al.(2025)Chatterji, Cunningham, Deming, Hitzig, Ong, Shan, and Wadman]{chatgpt2025}
A.~Chatterji, T.~Cunningham, D.~J. Deming, Z.~Hitzig, C.~Ong, C.~Y. Shan, and K.~Wadman.
\newblock {How People Use ChatGPT}, 2025.
\newblock URL \url{https://doi.org/10.3386/w34255}.

\bibitem[Corrigan-Gibbs and Boneh(2017)]{corrigan2017prio}
H.~Corrigan-Gibbs and D.~Boneh.
\newblock Prio: Private, robust, and scalable computation of aggregate statistics.
\newblock In \emph{14th USENIX symposium on networked systems design and implementation (NSDI 17)}, pages 259--282, 2017.

\bibitem[Durfee(2023)]{durfee2023unbounded}
D.~Durfee.
\newblock Unbounded differentially private quantile and maximum estimation.
\newblock \emph{Advances in Neural Information Processing Systems}, 36:\penalty0 77691--77712, 2023.

\bibitem[Dwork et~al.(2006)Dwork, McSherry, Nissim, and Smith]{DMNS06}
C.~Dwork, F.~McSherry, K.~Nissim, and A.~D. Smith.
\newblock Calibrating noise to sensitivity in private data analysis.
\newblock In S.~Halevi and T.~Rabin, editors, \emph{Theory of Cryptography, Third Theory of Cryptography Conference, {TCC} 2006, New York, NY, USA, March 4-7, 2006, Proceedings}, volume 3876 of \emph{Lecture Notes in Computer Science}, pages 265--284. Springer, 2006.
\newblock \doi{10.1007/11681878\_14}.
\newblock URL \url{https://doi.org/10.1007/11681878\_14}.

\bibitem[Eichner et~al.(2024)Eichner, Ramage, Bonawitz, Huba, Santoro, McLarnon, Van~Overveldt, Fallen, Kairouz, Cheu, et~al.]{eichner2024confidential}
H.~Eichner, D.~Ramage, K.~Bonawitz, D.~Huba, T.~Santoro, B.~McLarnon, T.~Van~Overveldt, N.~Fallen, P.~Kairouz, A.~Cheu, et~al.
\newblock Confidential federated computations.
\newblock \emph{arXiv preprint arXiv:2404.10764}, 2024.

\bibitem[Google(2020)]{ghs2020}
Google.
\newblock Advancing health research with google health studies.
\newblock \url{https://blog.google/technology/health/google-health-studies-app/}, 2020.

\bibitem[Google(2025{\natexlab{a}})]{cfc}
Google.
\newblock Confidential federated compute -- public tee-hosted binaries for verifiable server-side computation, 2025{\natexlab{a}}.
\newblock URL \url{https://github.com/google-parfait/confidential-federated-compute}.

\bibitem[Google(2025{\natexlab{b}})]{gemmacpp}
Google.
\newblock Gemma.cpp -- lightweight, standalone c++ inference engine for google's gemma models, 2025{\natexlab{b}}.
\newblock URL \url{https://github.com/google/gemma.cpp}.

\bibitem[Hu et~al.(2025)Hu, McKenna, Yu, Wu, Zhao, Xu, and Kairouz]{hu2025actgarldifferentiallyprivateconditional}
Y.~Hu, R.~McKenna, D.~Yu, S.~Wu, H.~Zhao, Z.~Xu, and P.~Kairouz.
\newblock {ACTG-ARL: Differentially Private Conditional Text Generation with RL-Boosted Control}, 2025.
\newblock URL \url{https://arxiv.org/abs/2510.18232}.

\bibitem[Huba et~al.(2022)Huba, Nguyen, Malik, Zhu, Rabbat, Yousefpour, Wu, Zhan, Ustinov, Srinivas, et~al.]{huba2022papaya}
D.~Huba, J.~Nguyen, K.~Malik, R.~Zhu, M.~Rabbat, A.~Yousefpour, C.-J. Wu, H.~Zhan, P.~Ustinov, H.~Srinivas, et~al.
\newblock Papaya: Practical, private, and scalable federated learning.
\newblock \emph{Proceedings of Machine Learning and Systems}, 4:\penalty0 814--832, 2022.

\bibitem[{Intel}(2025)]{intelTDX}
{Intel}.
\newblock Intel\textregistered {Trust Domain Extensions} ({Intel}\textregistered {TDX}), June 2025.
\newblock URL \url{https://cdrdv2.intel.com/v1/dl/getContent/690419}.
\newblock White Paper.

\bibitem[Kaggle(2025)]{kagglegemma}
Kaggle.
\newblock Gemma -- a family of lightweight, state-of-the-art open models from google, built from the same research and technology used to create the gemini models, 2025.
\newblock URL \url{https://www.kaggle.com/models/google/gemma-3}.

\bibitem[Kairouz et~al.(2021)Kairouz, Liu, and Steinke]{kairouz2021distributed}
P.~Kairouz, Z.~Liu, and T.~Steinke.
\newblock The distributed discrete gaussian mechanism for federated learning with secure aggregation.
\newblock In \emph{International Conference on Machine Learning}, pages 5201--5212. PMLR, 2021.

\bibitem[Liu et~al.(2025)Liu, Cohen, Ghazi, Kairouz, Kamath, Knop, Kumar, Manurangsi, Sealfon, Yu, and Zhang]{liu2025uraniadifferentiallyprivateinsights}
D.~Liu, E.~Cohen, B.~Ghazi, P.~Kairouz, P.~Kamath, A.~Knop, R.~Kumar, P.~Manurangsi, A.~Sealfon, D.~Yu, and C.~Zhang.
\newblock {Urania: Differentially Private Insights into AI Use}, 2025.
\newblock URL \url{https://arxiv.org/abs/2506.04681}.

\bibitem[Meta(2025)]{meta2025}
Meta.
\newblock Building private processing for {AI} tools on {WhatsApp}.
\newblock \url{https://engineering.fb.com/2025/04/29/security/whatsapp-private-processing-ai-tools}, 2025.

\bibitem[{NVIDIA}(2023)]{nvidiah100}
{NVIDIA}.
\newblock {NVIDIA H100 Tensor Core GPU Architecture}, 2023.
\newblock URL \url{https://resources.nvidia.com/en-us-hopper-architecture/nvidia-h100-tensor-c}.

\bibitem[Ongaro and Ousterhout(2014)]{raftusenix}
D.~Ongaro and J.~Ousterhout.
\newblock In search of an understandable consensus algorithm.
\newblock In \emph{2014 USENIX Annual Technical Conference (USENIX ATC 14)}, pages 305--319, Philadelphia, PA, June 2014. USENIX Association.
\newblock ISBN 978-1-931971-10-2.
\newblock URL \url{https://www.usenix.org/conference/atc14/technical-sessions/presentation/ongaro}.

\bibitem[Play(2025)]{recorder}
G.~Play.
\newblock {Recorder - Apps on Play Store}, 2025.
\newblock URL \url{https://play.google.com/store/apps/details?id=com.google.android.apps.recorder}.

\bibitem[{Ramage, Daniel and Mazzocchi, Stefano}(2020)]{google_federated_analytics_2020}
{Ramage, Daniel and Mazzocchi, Stefano}.
\newblock {Federated Analytics: Collaborative Data Science without Data Collection}, May 2020.
\newblock URL \url{https://research.google/blog/federated-analytics-collaborative-data-science-without-data-collection/}.
\newblock Google AI Blog.

\bibitem[Research(2024)]{apple2024}
A.~S. Research.
\newblock Private cloud compute: A new frontier for ai privacy in the cloud.
\newblock \url{https://security.apple.com/blog/private-cloud-compute}, 2024.

\bibitem[Srinivas et~al.(2025)Srinivas, Cormode, Honarkhah, Lurye, Hehir, He, Hong, Magdy, Huba, Wang, et~al.]{srinivas2025papaya}
H.~Srinivas, G.~Cormode, M.~Honarkhah, S.~Lurye, J.~Hehir, L.~He, G.~Hong, A.~Magdy, D.~Huba, K.~Wang, et~al.
\newblock $\{$PAPAYA$\}$ federated analytics stack: Engineering privacy, scalability and practicality.
\newblock In \emph{22nd USENIX Symposium on Networked Systems Design and Implementation (NSDI 25)}, pages 883--898, 2025.

\bibitem[Sun et~al.(2024)Sun, Kairouz, Sun, Gascon, and Suresh]{sun2024private}
Z.~Sun, P.~Kairouz, H.~Sun, A.~Gascon, and A.~T. Suresh.
\newblock Private federated discovery of out-of-vocabulary words for gboard.
\newblock \emph{arXiv preprint arXiv:2404.11607}, 2024.

\bibitem[Talwar et~al.(2024)Talwar, Wang, McMillan, Feldman, Bansal, Basile, Cahill, Chan, Chatzidakis, Chen, et~al.]{talwar2024samplable}
K.~Talwar, S.~Wang, A.~McMillan, V.~Feldman, P.~Bansal, B.~Basile, A.~Cahill, Y.~S. Chan, M.~Chatzidakis, J.~Chen, et~al.
\newblock Samplable anonymous aggregation for private federated data analysis.
\newblock In \emph{Proceedings of the 2024 on ACM SIGSAC Conference on Computer and Communications Security}, pages 2859--2873, 2024.

\bibitem[Tamkin et~al.(2024)Tamkin, McCain, Handa, Durmus, Lovitt, Rathi, Huang, Mountfield, Hong, Ritchie, Stern, Clarke, Goldberg, Sumers, Mueller, McEachen, Mitchell, Carter, Clark, Kaplan, and Ganguli]{clio2024}
A.~Tamkin, M.~McCain, K.~Handa, E.~Durmus, L.~Lovitt, A.~Rathi, S.~Huang, A.~Mountfield, J.~Hong, S.~Ritchie, M.~Stern, B.~Clarke, L.~Goldberg, T.~R. Sumers, J.~Mueller, W.~McEachen, W.~Mitchell, S.~Carter, J.~Clark, J.~Kaplan, and D.~Ganguli.
\newblock {Clio: Privacy-Preserving Insights into Real-World AI Use}, 2024.
\newblock URL \url{https://arxiv.org/abs/2412.13678}.

\bibitem[Team(2017)]{apple2017}
A.~P. Team.
\newblock Learning with privacy at scale.
\newblock \url{https://machinelearning.apple.com/2017/12/06/learning-with-privacy-at-scale.html}, 2017.

\end{thebibliography}

\appendix
\section{Access Policy}
\label{app:access_policy}
%\subsection{Example without autotuning}
To illustrate the space of permissible access policies, we begin with a policy that permits a differentially private group-by aggregation to be run, along with autotuning of its parameters.

\begin{lstlisting}[language=textproto]
pipelines {
  # Each logical pipeline name maps to
  # a LogicalPipelinePolicy.
  "name_of_logical_pipeline_policy": {
    # Each LogicalPipelinePolicy 
    # consists of PipelineVariantPolicy
    # instances.
    instances {
      transforms {
        src_node_ids: [0, 1]
        dst_node_ids: [1, 2]
        # Application binary reference
        # values.
        application {
          reference_values { ... }
        }
        # Application binary config 
        # specifying required privacy 
        # params for the DP group-by
        # aggregation with autotuning.
        config_constraints {
          ...
        }
      }
    }
  }
}
\end{lstlisting}

The \texttt{config\_constraints} field for this server-side workload should include the desired DP parameters $\varepsilon,\delta$ and specify that an upload can be used a maximum of one time in either a DP algorithm that computes DP quantiles or a DP algorithm that computes a group-by aggregation. The TEE binary that is permitted to execute the permitted computations is indicated by the \texttt{application}'s \texttt{reference\_values}. The lists \texttt{src\_node\_ids} and \texttt{dest\_node\_ids} reflect the distributed fashion in which we execute the DP computation: many leaf nodes (1) ingest the encrypted uploads (0), compute contribution-bounded noiseless intermediate aggregates in parallel---sums of slices of the input---then pass encrypted state to a root node (2) that merges and noises the aggregates.

By listing multiple pipeline variants in the policy, as below, we can decouple policy updates from rolling out new processing pipeline binary versions.

\begin{lstlisting}[language=textproto]
pipelines {
  # Each logical pipeline name maps to
  # a LogicalPipelinePolicy containing
  # one or more PipelineVariantPolicies.
  "name_of_logical_pipeline_policy": {
    # PipelineVariantPolicy admitting
    # binary v1
    instances {
      transforms {
        ...
      }
    }
    # PipelineVariantPolicy admitting
    # binary v2
    instances {
      transforms {
        ...
      }
    }
  }
}
\end{lstlisting}

Refer to the file \href{https://github.com/google-parfait/federated-compute/blob/main/fcp/protos/confidentialcompute/access_policy.proto}{\texttt{access\_policy.proto}} in the GitHub repository \href{https://github.com/google-parfait/federated-compute}{google-parfait/federated-compute} for more exposition on the policy structure.

\section{Autotuning}
\label{app:autotuning}
\subsection{Why Autotuning Requires DP Quantiles}
Suppose that we replaced the DP quantile algorithm in our autotuning system with a deterministic algorithm. We will demonstrate that such a variant is vulnerable to an attack that deduces which group a privacy unit contributes to, even when the tuned parameters are hidden from the attacker.

Consider the case where (a) the user's query requests a summation of one value, with \texttt{L\_inf}$=1$ but \texttt{max\_groups\_contributed} is not specified (b) Alice contributes to either the group ``normal'' or both groups ``normal'' and ``embarrassing.'' Because of (a), our system marks \texttt{max\_groups\_contributed} as the sole parameter to tune. This means it will add noise proportional to the $k=83$rd percentile of the number of groups that a privacy unit contributes to. If the adversary mixes Alice's (encrypted) input with 82 copies of ``normal'' privacy units and 17 copies of units that contribute to both ``normal'' and ``embarrassing'' then (b) implies that the number of groups Alice contributes to will be the 83rd percentile, which directly reveals which of the two inputs she has.

If the adversary directly observes the 83rd percentile, they can deduce whether she has contributed to ``embarrassing'' simply by mapping 2 to ``yes'' and 1 to ``no.'' If they do not, recall that the noise added to the sums is scaled proportional to the choice of \texttt{max\_groups\_contributed}; because the attacker has control over the Alice's peers, they can measure the noise via subtraction of pre-noise values and apply a likelihood test to determine which of the two distributions it was sampled from.

\subsection{Mathematical Analysis of Autotuning}
For simplicity, our analysis will make the assumption that we are tuning one parameter; to generalize to $g>1$ parameters, we can divide failure probabilities by $g$ and apply the union bound.

Let $X$ be the $n$ positive scalar values over which we are computing the $k$-th percentile; if we are tuning \texttt{L\_inf} for example then the $i$-th scalar is the maximum contribution to any group that the $i$-th privacy unit makes. Our autotuning process proceeds as follows:
\begin{enumerate}
    \item For each element of $X$, toss a coin with head-bias $q \ll 1/2$. Let $S$ denote the $m$ scalars whose coin landed heads; note that $m\sim \mathbf{Bin}(n,q)$.
    \item Feed $S$ into our $\varepsilon$-DP quantile algorithm to estimate the $k$-th percentile of $S$
    \item Report the outcome of that algorithm as an estimate of the $k$-th percentile of $X$.
\end{enumerate}

For any multiset $T$, we will say that we have a ``$\gamma$-approximation of the $k$-th percentile of $T$'' when its index in (sorted) $T$ lies inside the interval $G_\gamma(T) := [\frac{k-\gamma}{100}\cdot |T|, ~ \frac{k+\gamma}{100} \cdot |T|]$, the indices of values within $\gamma$\% of the $k$-th percentile. \emph{Our driving question is to find a setting of $q$ such that the outcome 10-approximates the $k$-th percentile of $X$}.

Observe that there are two sources of error: the sampling of $S$ and the DP algorithm. This appendix has one section for each. First, we find the range of $q$ such that $k$-th percentile of $S$ 5-approximates the $k$-th percentile of $X$, except with probability $\beta$. Then we find the range of $q$ such that the DP algorithm will report a 5-approximation of the $k$-th percentile of $S$, except with probability $2\beta$. Putting these two parts together with $\beta=1/30$ ensures that we have a range of $q$ such that the outcome 10-approximates the $k$-th percentile of $X$, 90\% of the time.

\subsubsection{Sampling rate to ensure representativeness}
Let $a$ be the number of elements in $S$ that are less than the $(k-5)$-th percentile of $X$.

Let $b$ be the number of elements in $S$ that lie between the $(k-5)$-th and $(k+5)$-th percentiles of $X$.

Let $c$ be the number of elements in $S$ that are greater than the $(k+5)$-th percentile of $X$.

If we could show $a < \frac{k}{100}|S|$ and $c < \frac{100-k}{100}|S|$, then the $k$-th percentile of $S$ approximates the $k$-th percentile of $X$: the first inequality implies that the $k$-th percentile of $S$ is larger than the $(k-5)$-th percentile of $X$ and the second implies it is smaller than the $(k+5)$-th percentile of $X$.

Because $|S| = a+b+c$, we can rewrite our desired constraints:
\begin{align*}
a <{}& \frac{k}{100}|S|\\
    ={}& \frac{k}{100}(a+b+c)\\
\frac{100-k}{100}a <{}& \frac{k}{100}(b+c)\\
\frac{100-k}{k}a <{}& b+c \stepcounter{equation} \tag{\theequation} \label{eq:a-constraint}
\end{align*}
and
\begin{align*}
c <{}& \frac{100-k}{100}|S|\\
    ={}& \frac{100-k}{100}(a+b+c)\\
\frac{k}{100}c <{}& \frac{100-k}{100}(a+b)\\
\frac{k}{100-k}c <{}& a+b \stepcounter{equation} \tag{\theequation} \label{eq:c-constraint}
\end{align*}

Observe that the \emph{marginal} distributions are $a\sim \mathbf{Bin}(\frac{k-5}{100}n, q)$, $b\sim \mathbf{Bin}(\frac{1}{10}n,q)$, and $c\sim \mathbf{Bin}(\frac{95-k}{100}n,q)$. We will use a concentration inequality and a union bound to derive constraints on $q$ such that we can derive \eqref{eq:a-constraint} and \eqref{eq:c-constraint}.

\newcommand{\paren}[1]{\ensuremath{\left( {#1} \right)}}
\newcommand{\pr}[2]{{\ifx&#1& \mathbb{P} \else \underset{#1}{\mathbb{P}} \fi \left[#2\right]}}
\newcommand{\abs}[1]{\left| #1 \right|}

Suppose $\eta$ is a binomial with mean $\mu$. For any $\psi \in(0,1)$, the multiplicative Chernoff bound\footnote{It is also possible to use the additive Hoeffding bound, but recall we target $q\ll 1/2$ in which case the multiplicative bound is tighter.} implies the two inequalities below:
\begin{align*}
\pr{}{|\eta - \mu| \geq \psi\mu} &\leq 2\exp(-\psi^2\mu/3)\\
\pr{}{\eta \leq (1-\psi)\mu } &\leq \exp(-\psi^2\mu/2)
\end{align*}

We apply these tail bounds to the marginals of $a,b,c$:
\begin{align*}
&\pr{}{\abs{a-\frac{k-5}{100}nq} \geq \psi \cdot \frac{k-5}{100}nq}\\
&~~~\leq 2\exp\paren{-\psi^2\paren{\frac{k-5}{100}nq}/3}\\
&\pr{}{b \leq (1-\psi) \cdot \frac{1}{10}nq} \leq \exp\paren{-\psi^2\paren{\frac{1}{10}nq}/2}\\
&\pr{}{\abs{c-\frac{95-k}{100}nq} \geq \psi \cdot \frac{95-k}{100}nq}\\
&~~~\leq 2\exp\paren{-\psi^2\paren{\frac{95-k}{100}nq}/3}
\end{align*}
A union bound implies that we can safely assume all three of the inequalities
\begin{align*}
    \abs{a-\frac{k-5}{100}nq} &< \psi \cdot \frac{k-5}{100}nq\\
    b &> (1-\psi) \cdot \frac{1}{10}nq\\
    \abs{c-\frac{95-k}{100}nq} &< \psi \cdot \frac{95-k}{100}nq
\end{align*}
are simultaneously true, except with probability
$$
\leq 5\exp\paren{-\psi^2\cdot \min \paren{ \frac{k-5}{300}, \frac{1}{20}, \frac{95-k}{300}} \cdot nq}
$$
So to target a failure probability of $\beta$, we should ensure \begin{equation*}\boxed{q \geq \frac{1}{\psi^2 n}\cdot \max\paren{\frac{300}{k-5}, 20, \frac{300}{95-k}} \cdot \ln \frac{5}{\beta}.}\end{equation*}
Once we upper bound $\psi$, we will have our desired result. We do so by  combining \eqref{eq:a-constraint} and \eqref{eq:c-constraint} with the concentration bounds. To ensure \eqref{eq:a-constraint}, it will suffice to show that an upper bound on $\frac{100-k}{k}a$ is less than a lower bound on $b+c$:
\begin{align*}
    \frac{100-k}{k}a &< \frac{100-k}{k}\cdot (1+\psi) \cdot \frac{k-5}{100} \cdot nq\\
    b+c &> \frac{105-k}{100}\cdot (1-\psi)\cdot nq
\end{align*}
So it suffices for $\frac{100-k}{k}\cdot (1+\psi) \cdot \frac{k-5}{100} \cdot nq < \frac{105-k}{100}\cdot (1-\psi)\cdot nq$ to imply \eqref{eq:a-constraint}:
\begin{align*}
    \frac{100-k}{k}\cdot (1+\psi) \cdot \frac{k-5}{100} \cdot nq &< \frac{105-k}{100}\cdot (1-\psi)\cdot nq\\
    \frac{1+\psi}{1-\psi} &< \frac{k(105-k)}{(k-5)(100-k)}
\end{align*}

To ensure \eqref{eq:c-constraint}, it will suffice to show that an upper bound on $\frac{k}{100-k}c$ is less than a lower bound on $a+b$:
\begin{align*}
    \frac{k}{100-k}c &< \frac{k}{100-k}\cdot (1+\psi)\cdot \frac{95-k}{100}\cdot nq\\
    a+b &> (1-\psi)\cdot \frac{k+5}{100}\cdot nq
\end{align*}
So it suffices for $\frac{k}{100-k}\cdot (1+\psi)\cdot \frac{95-k}{100}\cdot nq < (1-\psi)\cdot \frac{k+5}{100}\cdot nq$ to imply \eqref{eq:c-constraint}:
\begin{align*}
    \frac{k}{100-k}\cdot (1+\psi)\cdot \frac{95-k}{100}\cdot nq &< (1-\psi)\cdot \frac{k+5}{100}\cdot nq\\
    \frac{1+\psi}{1-\psi} &< \frac{(100-k)(k+5)}{k(95-k)}
\end{align*}

Hence, we should set $\psi$ such that
$$
\frac{1+\psi}{1-\psi} < \underbrace{\min \paren{\frac{k(105-k)}{(k-5)(100-k)}, \frac{(100-k)(k+5)}{k(95-k)}}}_z
$$
This simplifies to the inequality $\boxed{\psi < \frac{z-1}{z+1}}$; in practice, we search for this value numerically

\subsubsection{Sampling rate to ensure accurate DP quantile estimation}
We first find an integer $m^*$ such that when the sample size $m$ exceeds $m^*$, our DP algorithm will 5-approximate the $k$-th percentile of $S$ except with probability $\beta$. Then we find a range of $q$ such that $m>m^*$ except with probability $\beta$.

\paragraph{Size of $S$ to ensure accurate DP quantile estimation}

We begin with a brief sketch of the quantile-finding algorithm. It implements a simple variant of the \texttt{AboveThreshold} algorithm. The first idea is to divide the non-negative reals into a grid, where the grid points first follow linear growth then transition to exponential growth. At each point on this grid, the algorithm makes a noisy measurement of the number of inputs that are to the left. The output is the first grid point where this measurement exceeds a noisy threshold.

A bit more precisely,
\begin{enumerate}
    \item Initialize $p\gets 0.1$
    \item Sample $\eta\sim $ a Laplace distribution with scale $2/\eps$
    \item While $p$ has not crossed the maximum representable value,
    \begin{enumerate}
        \item Sample $\eta_p \sim $ a Laplace distribution with scale $4/\eps$
        \item Let $f_S(p)$ be the number of inputs that are at most $p$
        \item If $f_S(p)+\eta_p \geq \frac{k}{100}\cdot m + \eta$, break
        \item Else, $p\gets \max(p+0.1,  p \cdot 1.01 )$
    \end{enumerate}
    \item Return $p$
\end{enumerate}
Refer to the \texttt{DPQuantileAggregator} class in our library for the complete code.

Assuming the maximum is $2^{64}$, the number of Laplace random variables is $\leq 1 + 10 + 64 \log_{1.01} 2 < 4470$.

Except with probability $\beta$, all Laplace random variables have magnitude $< \frac{4}{\eps}\ln \frac{4470}{\beta}$. This means the algorithm will never halt on some $p$ where $\frac{k}{100}\cdot m - f_S(p) > \frac{8}{\eps}\ln \frac{4470}{\beta}$. By the same logic, it will never continue once it has found some $p$ where $f_S(p) - \frac{k}{100}\cdot m > \frac{8}{\eps}\ln \frac{4470}{\beta}$.

The upshot is that when $\boxed{m\geq m^* := \frac{160}{\eps}\ln \frac{4470}{\beta}}$, $\beta$ bounds the probability that the returned $p$ is not an approximation of a value inside of $G_5(S)$. The error of approximation is either an additive $0.1$ or a relative $\leq 1\%$.

\paragraph{Sampling rate to ensure minimum size of $S$}
Here, we choose $q$ such that the random variable $|S|$ is above the requisite threshold $m^*$ except with probability $\beta$.

If we want $\pr{}{|S| \leq m^*} \leq \beta$ and the multiplicative Chernoff bound implies $\pr{}{|S| \leq (1-\psi) nq} \leq \exp(-\psi^2 nq / 2)$, it will suffice to show
\begin{enumerate}[label=(\alph*)]
    \item $\pr{}{|S| \leq m^*} \leq \pr{}{|S| \leq (1-\psi)nq}$ and
    \item $\exp(-\psi^2 nq / 2) \leq \beta$.
\end{enumerate}

The constraint (a) is equivalent to $m^*\leq (1-\psi)nq$ which is the same as $q \geq \frac{m^*}{(1-\psi)n}$. Meanwhile, constraint (b) is equivalent to $q \geq 2\ln(1/\beta) / n\psi^2$. Hence, it suffices for
\begin{equation*}
\boxed{q \geq \max\paren{ \frac{2}{n\psi^2}\ln\frac{1}{\beta}, \frac{m^*}{(1-\psi)n}}}
\end{equation*}

It remains to find $\psi$ that minimizes the above. The first function is a decreasing function of $\psi$ while the former is an increasing function, which means the minimum of the maximum is achieved where they intersect. In turn, we just need to \underline{solve the quadratic} $\boxed{\psi^2 = (1-\psi) \frac{2}{m^*} \ln(1/\beta)}$; in practice, we search for this value numerically.

\end{document}